\documentstyle [12pt,twoside]{article}

\newcommand{\bq}{\begin{equation}}
\newcommand{\ba}{\begin{eqnarray}}
\newcommand{\eq}{\end{equation}}
\newcommand{\ea}{\end{eqnarray}}

\def\bo{{\raise.15ex\hbox{\large$\Box$}}}
\def\bob{{\lower.2ex\hbox{\large$\Box$}}}

\def\TH{{\raise.2ex\hbox{$\displaystyle \bigodot$}\mskip-4.7mu \llap H \;}}

\catcode`@=11
\def\underline#1{\relax\ifmmode\@@underline#1\else
        $\@@underline{\hbox{#1}}$\relax\fi}
\catcode`@=12

\begin{document}

\hfill LA-UR-95-2091
\centerline{\large{\bf Classical Dynamics for Linear Systems: The Case
of}}
\centerline{\large{\bf Quantum Brownian Motion}}

\vspace{2cm}

\centerline{\bf James Anglin$^\star$ and Salman Habib$^\dagger$}

\vspace{1cm}

\centerline{\em Theoretical Division}
\centerline{\em Mail Stop B288}
\centerline{\em Los Alamos National Laboratory}
\centerline{\em Los Alamos, NM 87545}

\vspace{2cm}

\centerline{\bf Abstract}

It has long been recognized that the dynamics of linear quantum systems
is classical in the Wigner representation. Yet many conceptually
important linear problems are typically analyzed using such generally
applicable techniques as influence functionals and Bogoliubov
transformations.  In this Letter we point out that the classical
equations of motion provide a simpler and more intuitive formalism for
linear quantum systems. We examine the important problem of Brownian
motion in the independent oscillator model, and show that the quantum
dynamics is described directly and completely by a c-number Langevin
equation.  We are also able to apply recent insights into quantum
Brownian motion to show that the classical Fokker-Planck equation is
always local in time, regardless of the spectral density of the
environment.
\vfill
\noindent e-mail:\\
\noindent $^\star$anglin@t6-serv.lanl.gov\\
\noindent $^\dagger$habib@predator.lanl.gov\\
\newpage

It has long been recognized that, although the Wigner function of a
linear quantum mechanical system cannot in general be considered as a
classical ensemble density, yet it evolves according to exactly the
same Liouville equation as the corresponding classical system: $\dot f
= \hat L_{cl} f$.\cite{fdot} This is true even for linear systems
with time-dependent parameters, and for non-linear systems treated in
Gaussian approximation \cite{gauss}. Nevertheless, conceptually
important problems, such as quantum Brownian motion and particle
production in classical background fields, are generally treated
using formalisms which do not explicitly take advantage of
classicality, such as path integrals and Bogoliubov transformations.
An important example of the disadvantages of this unnecessary
sophistication is the way in which the machinery of influence
functionals has successfully hidden the fact that, in the independent
oscillator model \cite{io}, quantum and classical Brownian motion are
dynamically {\em identical}.

The more sophisticated approaches are often technically cumbersome,
and their interpretation frequently seems rather subtle. It is usually
a challenge, for example, to associate a simple physical intuition
with the closed time paths of an influence functional. In contrast, a
much more direct approach to any linear problem is to use the
classical equations of motion to obtain the time evolution of the
Wigner function. After solving these equations, with their (usually
easiest) Cauchy boundary conditions, one has nothing further to do
(unless one is studying an open system, in which case there remains
the integration over the initial data of the environment). And since
the classical trajectories of a system are easy to associate with a
consistent physical intuition, one can not only compute the correct
final answer, but also appreciate the intermediate processes that
generate the ultimate result. Since most linear problems are
essentially of conceptual interest, this is certainly a consummation
to be wished: Classical mechanics can greatly simplify quantum
mechanics.

In the linear regime, the converse can also be true. It is widely
assumed that, even with the simple environmental model of independent
oscillators, the classical Fokker-Planck equation for a Brownian
oscillator is only local in time if one assumes a particular spectral
density for the environmental bath. This is not true. By applying an
insight from a recent solution of the quantum problem, we can abandon
the Markovian assumption in classical Brownian motion, and still
obtain a local Fokker-Planck equation. In the non-Markovian regime,
the correct identification of the physical noise and the effective
application of the fluctuation-dissipation relation become much more
complicated \cite{ngvk/rz}. Without the use of methods inspired by
quantum mechanics, it would be very difficult indeed to discern the
Fokker-Planck equation: In this problem, quantum mechanics can greatly
simplify classical mechanics. Moreover, the demonstration that the
classical problem is indeed dynamically identical to the quantum
problem must remove any lingering suspicion that influence functionals
(or any other essentially quantum techniques) are necessary to capture
the quantum dynamics.

To proceed to explicit details, we recall the Hamiltonian for the
Brownian harmonic oscillator linearly coupled to a bath of independent
oscillators:
\begin{equation}\label{Ham}
H={1\over 2m}P^2 + {m\over 2}\Omega^2Q^2 +
{1\over2}\int_0^\infty\!d\omega\, \Bigl[p_\omega^2 + (\omega q_\omega
- g_\omega x)^2\Bigr]~,
\end{equation}
where $g_\omega$ is the coupling constant, whose $\omega$-dependence
effectively incorporates the spectral density of the environment. We
also recall the Liouville propagator for the probability density in
phase space:
\begin{eqnarray}\label{fprop}
&& f(Q_F,P_F;\{q_{\omega F},p_{\omega F}\};t) = \int\! dQ_I dP_I Dq_I
Dp_I\, \delta\bigl(P_F-P(t)\bigr)\delta\bigl(Q_F-Q(t)\bigr)\nonumber\\
&&\qquad\qquad\qquad\qquad\qquad\qquad\times\int\prod_{\omega}
	\delta\bigl(q_{\omega F} - q_\omega(t)\bigr)
	\delta\bigl(p_{\omega F} - p_\omega(t)\bigr)\nonumber\\
&&\qquad\qquad\qquad\qquad\qquad\qquad\qquad\times
	f(Q_I,P_I;\{q_{\omega I},p_{\omega I}\},t_I)\;,
\end{eqnarray}
where $Q(t),q_\omega(t)$ and $P(t),p_\omega(t)$ denote the values of
the canonical variables at time $t$ after evolution, under Hamilton's
equations, from the initial values $Q_I,P_I$ and $\{q_{\omega
I},p_{\omega I}\}$ at time $t_I$.  We write $Q(t)$ instead of $Q(t;
Q_I, P_I, \{q_{\omega I}, p_{\omega I}\})$, {\em etc.}, only in order
to avoid an unwieldy notation.  It will be important to remember,
despite this shorthand convention, that $Q(t)$ and $P(t)$ are
time-dependent linear combinations of all the initial variables.

As in the standard theoretical version of Brownian motion, we assume
that the intitial ensemble factorizes and that its environmental part
is Gaussian\footnote{Slight variations in this initial distribution may
be contemplated, as may variations in the Hamiltonian that are
obtained by canonical transformations mixing the system and
environmental sectors. These variations can be encompassed by
straightforward generalizations of the present discussion.}:
\begin{equation}\label{barf}
f(Q_I,P_I;\{q_I(\omega),p_I(\omega)\},t_I) =
\bar{f}(Q_I,P_I;t_I)\,\prod_{\omega}\Bigl
({\omega\beta_\omega\over2\pi}\Bigr)
e^{-{1\over2}\beta_\omega[\omega^2 q_{\omega I}^2 + p_{\omega
I}^2]}\;.
\end{equation}
Since we wish to know only the reduced distribution $\bar{f}(Q,P;t)$
over the observed sector of phase space, we integrate over the
environmental sector in (\ref{fprop}). With the factorized initial
condition (\ref{barf}), this is equivalent to propagating $\bar f$
using the equations of motion $P(t) = m\dot{Q}(t)$ and
\begin{equation}\label{Langevin}
\ddot{Q}(t)+\Omega^2Q(t) + K(t)Q_I + \int_0^t\!dt'K(t-t')\dot{Q}(t') =
{F(t)\over m}\,,
\end{equation}
where $K(t)\equiv (1/m)\int_0^\infty\!d\omega\,g_\omega^2\cos\omega
t$, and $F(t)$ is a stochastic force. Since the environmental part of
(\ref{barf}) is Gaussian, $F$ is completely described by its two-point
correlation function:
\begin{equation}\label{FDR}
\langle F(t) F(t')\rangle = \int_0^\infty\!d\omega\,
{g_\omega^2\over\beta_\omega}\cos\omega(t-t')\;.
\end{equation}

Eqn. (\ref{Langevin}) is the (generalized) Langevin equation.  The
standard classical approach to this problem relies on separating the
systematic and stochastic terms in this equation \cite{ngvk/rz}. In the
special case where $\pi g_\omega^2 \equiv 4m\gamma$ is constant, the
integral kernel $K$ becomes a delta function, and one obtains a
stochastic differential equation with Ohmic damping.  The entire LHS of
(\ref{Langevin}) is then systematic, and the RHS can be unambiguously
identified as the noise. The usual classical prescription $\beta_\omega
= (k_{B}T)^{-1}$, independent of $\omega$ as well, also ensures that
the environmental noise is white. In this case, it is easy to use the
fluctuation-dissipation relation $\langle F(t)F(t')\rangle = 4m\gamma
k_BT \delta(t-t')$ to derive an evolution equation for the reduced
phase space distribution $\bar f$, which is local in time: the
Fokker-Planck equation.

For general $g_\omega$, however, Langevin equations are
integro-differential, incorporating a back-reaction term with memory.
The noise is also colored in general ({\it i.e.}, the correlation
function is not a delta function), even with constant $\beta_\omega$.
One might think that these circumstances would prevent the generalized
Fokker-Planck equation from being local. In fact, because the memory
term depends on the past history of $\dot{Q}$, over all of which it
has been affected by the environmental noise, it actually contains an
implicit stochastic part. So the physical noise receives a
contribution from back-reaction, as well as the explicit ``bare
noise.'' The crucial identification of the systematic and stochastic
parts of the Langevin equation is therefore more subtle.

One can still derive a generalized Fokker-Planck equation by the
standard method, using the colored physical noise that includes
back-reaction, and the non-local systematic evolution. Remarkably, it
will turn out that a rather complicated kind of
fluctuation-dissipation relation can be found, which will guarantee
that the generalized Fokker-Planck equation derived from Eqn.
(\ref{Langevin}) will always be local in time \cite{jash}. But this
approach is quite cumbersome.

A much shorter derivation is available, inspired by the quantum
mechanical treatment of Ref. \cite{Paz}. Integrating over the
environmental sector in Eqn. (\ref{fprop}), and using a
representation for the two surviving delta functions that is more
familiar in quantum contexts, we can write
\begin{eqnarray}\label{fdot}
&&\dot{\!\!\bar f}(Q_F,P_F;t) =
i\int\!dQ_IdP_I\,\bar{f}(Q_I,P_I;t_I)\int\!Ddq_I Dp_I
{}~\hbox{e}^{-{1\over2}\int_0^\infty\!d\omega\,\beta_\omega[\omega^2
q_{\omega I}^2 + p_{\omega I}^2]}\nonumber\\
&&\quad \times \int\!{dk dk'\over (2\pi)^2}\,
 e^{ik[Q_F - Q(t)]+k'[P_F - P(t)]}
  \Bigl(k \dot{Q}(t) - k'\dot{P}(t)\Bigr)\;.\nonumber\\
\end{eqnarray}
Because the Langevin equation is linear, the solutions $Q(t), P(t)$
depend linearly on the initial variables $\{q_{\omega I},p_{\omega
I}\}$ and $Q_I,P_I$.

The $k$ and $k'$ pre-factors in (\ref{fdot}) can be replaced by
derivatives under the integral with respect to $Q_F$ and $P_F$. The
thus-restored delta functions can then be used to replace the $Q_I$ and
$P_I$ terms in $\dot{Q}(t)$ and $\dot{P}(t)$ with linear combinations
of $Q_F$, $P_F$, and the $q_{\omega I}$ and $p_{\omega I}$.  The
Gaussian integrals over the environmental variables can then be
performed. The prefactors proportional to $q_{\omega I}$ and
$p_{\omega I}$ lead to terms proportional to $k$ and $k'$, which can
once again be expressed as derivatives with respect to $Q_F$ and
$P_F$. At the end of this process, we drop the subscripts on $Q_F$ and
$P_F$, and obtain an equation of the form
\begin{equation}\label{fdot2}
\dot{\!\!\bar f} = -{P\over m} {\partial\bar f\over\partial Q}
	+ m\bar\Omega^2(t) Q {\partial\bar f\over\partial P}
	+ 2\bar\gamma(t){\partial\ \over\partial P}P\bar{f}
        + d(t){\partial^2\bar{f}\over\partial Q\partial P}
	+ D(t){\partial^2\bar{f}\over\partial P^2}   \;.
\end{equation}
The time-dependent co-efficients may be expressed in terms of the
parameters in the Hamiltonian (\ref{Ham}). Their calculation is
straightforward but tedious, and we will not perform it here.  Instead
we merely note that time-dependent co-efficients, as well as the
so-called ``anomalous diffusion'' term proportional to $d(t)$, both
previously encountered in quantum treatments, do also appear in the
classical problem.

If we set $\beta_n = {2\over\hbar\omega_n}\tanh{\hbar\omega_n\over2k_B
T}$ in (\ref{barf}) and (\ref{fdot}), then equation (\ref{fdot2}) is
the Hu-Paz-Zhang master equation \cite{hpz}, in the Wigner
representation. Any positive choice of $\beta_n$, however, will still
provide a local equation for $\dot{\!\!\bar f}$.  And while $\beta_n =
(k_B T)^{-1}$ is prescribed by classical statistical mechanics (and
for Ohmic spectra this differs from quantum mechanics in permitting
white noise for all temperatures), any positive $\beta_n$ still
defines an allowed ({\it i.e.}, positive) classical ensemble density
for the environment. In the case of factorized, environmentally
Gaussian initial conditions, then, classical and quantum Brownian
motion can be said to differ only in the constraints prescribed for
$\bar{f}(Q_I,P_I;t_I)$. Dynamically, the quantum and classical cases
are identical.

Moving beyond the specific linear problem of Brownian motion, it is
straightforward to verify that, for the most general time-dependent
quadratic Hamiltonian, the propagator for the quantum mechanical
Wigner function is of exactly the same form as Eqn. (\ref{fprop}).
This means that the quantum Liouville equation for a linear system in
the Wigner representation will be the same as the classical Liouville
equation for the corresponding classical Hamiltonian.  Moreover, this
means that {\em the classical equations of motion}, with their
classically causal boundary conditions, {\em provide as exact and
complete an expression of linear quantum dynamics as any operator
equation or path integral}. And for linear open quantum systems with
an initially Gaussian environment, the reduced dynamics is {\em
directly} expressed in the c-number Langevin equations.

These c-number equations are both conceptually and computationally much
less demanding than the more general formulations of quantum dynamics.
If one wishes to obtain the final Wigner function $f(Q,P;t)$ of a
linear quantum system, one need only solve the time-reversed classical
equations of motion to determine the initial phase-space point
corresponding to the final point $(Q,P)$.  The initial Wigner function
at this initial point is the quantity sought. The Green's functions
needed to do this for final Cauchy conditions are typically much easier
to use than those required for the mixed initial and final conditions
of other approaches, and since the initial Wigner function is given,
one need do no further work once the classical equations are solved.
And of course this program remains effective for linear systems with
any number of degrees of freedom.

Furthermore, in many cases one may be able to gain considerable
information from the equations of motion alone, with only qualitative
discussion of the initial Wigner function. For example, knowing that a
bath of independent oscillators provides a Langevin noise, one can
conclude that environmental fluctuations will tend to even out
oscillations in the Wigner function of a Brownian oscillator. Rapid
oscillations between positive and negative values will be suppressed
quickly, and this provides one of the simplest intuitive pictures of
the process of decoherence (since such oscillations are the Wigner
signature of interference between classically distinct pointer
states).

As a second and less trivial example, we can also use classical
Langevin equations to understand why so-called ``supra-Ohmic''
environments tend to induce little decoherence in Brownian oscillators
\cite{supra} (a more complete discussion will be given elsewhere
\cite{jash}). In the present formalism, supra-Ohmic environments are
those for which $g_\omega^2$ increases with $\omega$ (up to some
cut-off scale, generally taken to be much larger than all other scales
in the problem).  If one compares Ohmic and supra-Ohmic models in
which $g_\Omega^2$ are equal, one typically finds that (after a few
cut-off timescales) the supra-Ohmic environment produces weaker
diffusion in the Brownian oscillator.  This is because stronger
coupling at high frequencies means that adiabatic dragging of fast
environmental oscillators gives the Brownian particle a higher
effective mass. Indeed, Brownian motion in supra-Ohmic environments
heavily weighted in the ultraviolet rapidly becomes indistinguishable
from Ohmic motion with a renormalized mass \cite{jash}.

One can easily anticipate this effect by using an adiabatic
approximation to write an effective Lagrangian without the fast degrees
of freedom, but it is interesting to see how it appears in the exact
solution of the full problem.  Computing the co-efficients in the
master equation (\ref{fdot2}) merely confirms that diffusion dies away;
studying the Langevin equation (\ref{Langevin}) reveals how this
occurs. After solving the equation, one can always in principle
re-write it in the form
\begin{equation}\label{Langloc}
\ddot{Q}(t)+\tilde\Omega^2(t)Q(t) + 2\gamma (t)\dot{Q}(t) = {1\over m}
[F(t) + F_{BR}(t)]\ ,
\end{equation}
where in addition to the bare force $F$ (whether stochastic or
systematic) there also appears the back-reaction force $F_{BR}$.  In
the case of a UV-dominated supra-Ohmic environment, there is a brief
inertial epoch that lasts for a few cut-off times, during which the
system behaves much as an Ohmic model, and the back-reaction force is
small. After this epoch, however, a remarkable regime emerges in which
any impulse applied to the Brownian oscillator as a bare force is
rapidly echoed in the back-reaction force, with opposite sign and
nearly equal magnitude \cite{jash}.  This ``counter-punch'' effectively
suppresses all forces acting on the oscillator, unless they vary
rapidly on the cut-off timescale, by a large factor equal to the ratio
of bare and renormalized masses found in the adiabatic analysis.  Thus
we achieve a nice intuitive picture, in which the supra-Ohmic
environment appears to deliberately isolate the Brownian particle, and
in which the importance of back-reaction is clearly exhibited.

When using classical intuition in quantum problems, one must of course
be careful. Wigner functions cannot, in general, be interpreted as
probability densities, because they can be negative. Even when one
happens to have a positive definite Wigner function, as initially one
has in the environmental sector for most Brownian motion problems, one
cannot assume that the classical interpretation will persist for all
times. For example, the picture just presented of decoherence, as due
to classically stochastic environmental noise, may seem to disagree
with the usual explanation in purely quantum language, according to
which the different branches of a ``Schr\"odinger's Cat state'' excite
orthogonal states of the environment \cite{whz}.

Actually, the two explanations can be reconciled by realizing that our
classical description effectively used a mixture of the Schr\"odinger
picture for the Brownian system and the Heisenberg picture for the
environment. It was thus able to refer only to the initial Wigner
function of the environment, which was interpretable as a probability
density. But if we were to use a uniformly Schr\"odinger picture, and
follow the evolution of the entire Wigner function before integrating
over the environmental sector, we would see that the negative values
in the initial Wigner function of the system spread rapidly into the
environmental sector. When the environmental integrals are performed,
these contaminating oscillations into negative values provide
cancellations that are the exact analogue of the vanishing inner
products in the standard discussion. So the fact that the environment
can affect the observed system only as a stochastic force does not
allow us to conclude that the environment actually remains in a state
described by a probabilistic ensemble.

Even if a Wigner function is positive definite, it still obeys a
non-classical constraint associated with the Uncertainty Principle:
\begin{equation}\label{UP}
\int\!d\Gamma\,f^2 \leq 2\pi\hbar\;,
\end{equation}
where $\int\!d\Gamma$ denotes the integral over all of phase space.
This constraint has important consequences. For example, a positive
definite Wigner function has a well-defined Boltzman entropy, but this
entropy will generally be larger than the von Neumann entropy. This
can be interpreted as being due to the fact that nearby points in
phase space are not distinct events in quantum mechanics, but are in a
sense parts of the same object. So the von Neumann entropy represents
a special kind of coarse-grained entropy.

(This idea provides a way to express the distinct quantum mechanical
concepts of correlation and entanglement, in the Wigner
representation. Compare the following two Wigner functions, each for
a system of two oscillators:
\begin{eqnarray}\label{2Wig}
f_1 &=& Z_1 e^{-{1\over m\Omega\hbar}(P_1^2 + P_2^2)}
[e^{-{m\Omega\over\hbar}(Q_1-a)^2}
e^{-{m\Omega\over\hbar}(Q_2-a)^2}\nonumber\\
& &\qquad\qquad\qquad\qquad \;\mbox{} + e^{-{m\Omega\over\hbar}(Q_1+a)^2}
e^{-{m\Omega\over\hbar}(Q_1+a)^2}]\nonumber\\
f_2 &=& Z_2 e^{-{1\over m\Omega\hbar}(P_1^2 + P_2^2)}
e^{-{m\bar\Omega\over\hbar}(Q_1^2 + Q_2^2)} e^{\gamma Q_1 Q_2}\;.
\end{eqnarray}
Classically, one would say that both of these functions describe
correlations between oscillators 1 and 2. But quantum mechanically,
the first describes a correlation (for $a^2>>{\hbar\over m\Omega}$),
while the second describes an entanglement. We can conclude that
entanglement is a correlation between parts of a single quantum
mechanical object \cite{ap}, and that the distinction between
correlation and entanglement is not a separate piece of ``quantum
magic'' from the uncertainty relation, but is actually a consequence
of it.)

Despite the subtleties involved in interpreting Wigner functions, we
emphasize that as long as one only uses classical intuition to
understand how the classical equations of motion propagate them from
initial to final times, there is nothing that can go wrong. This
careful use of classical intuition can be of considerable value in
appreciating the physics of linear quantum systems. One can expect to
shed new light on the meaning of some important quantum mechanical
effects, by identifying aspects of classical dynamics that are
responsible for their generation. This is the good news, but it has a
more pessimistic complement.

Because illustrations in linear models of basic quantum concepts are
usually treated using techniques applicable to non-linear
models as well, it is easy to receive an impression that these models
are providing an insight into essentially quantum physics. It is
important to realize, however, that the only specifically quantum
aspects of any of these models reside in the initial conditions. To
probe genuine quantum dynamics, we must accept the challenge of
non-linearity, and must not confuse generality of formalism with
generality of physics.

For their encouragement, and sharing of many insights, we would like
to thank Bei-lok Hu (JA and SH), Emil Mottola, Bill Unruh, and Robert
Zwanzig (SH). The support of the Natural Sciences and Engineering
Research Council of Canada (JA) and the United States Department of
Energy (SH) is gratefully acknowledged.


\begin{thebibliography}{99}
\bibitem{fdot} {See, {\em e.g.}, M. Hillery, R. O'Connell,
M. O. Scully, and E. P. Wigner, {\em Phys. Rep.} {\bf 106}, 121
(1984); V. I. Tatarskii, {\em Usp. Fiz. Nauk} {\bf 139}, 587 (1983)
(Sov. Phys. Usp. {\bf 26}, 311 (1983)).}
\bibitem{gauss} S. Habib, F. Cooper, and E. Mottola (in preparation).
\bibitem{io} R. J. Rubin, {\em J. Math. Phys.} {\bf 1}, 309 (1960);
{\em ibid} {\bf 2}, 373 (1961); G. W. Ford, M. Kac, and P. Mazur, {\em
J. Math. Phys.} {\bf 6}, 504 (1965); A. O. Caldeira and A. J. Leggett,
{\em Physica} {\bf 121A}, 587 (1983); H. Grabert, P. Schramm, and
G.-L. Ingold, {\em Phys. Rep.} {\bf 168}, 115 (1988).
\bibitem{ngvk/rz} For discussions of the subtleties in defining
``noise'' in Langevin equations, see N. G. van Kampen, {\em Stochastic
Processes in Physics and Chemistry} (North-Holland, Amsterdam, 1981);
R. Zwanzig, in {\em Systems Far From Equilibrium}, edited by L. Garrido
(Springer-Verlag, Berlin, 1980).
\bibitem{jash} J. R. Anglin and S. Habib, (in preparation).
\bibitem{Paz} J. P. Paz, in {\em The Physical Origin of Time Asymmetry},
edited by J. J. Halliwell {\em et al} (Cambridge University Press,
Cambridge, 1994).
\bibitem{hpz} B. L. Hu, J. P. Paz, and Y. Zhang, {\em Phys. Rev. D}
{\bf 45}, 2843 (1992).
\bibitem{supra} P. M. V. B. Barone and A. O. Caldeira, {\em
Phys. Rev. A} {\bf 43}, 57 (1991); J. P. Paz, S. Habib, and
W. H. Zurek, {\em Phys. Rev. D} {\bf 47}, 488 (1993).
\bibitem{whz} See, {\em e.g.}, W. H. Zurek, {\em Phys. Today} {\bf
44}, No. 10, 36 (1991).
\bibitem{ap} A. Peres, {\em Quantum Theory: Concepts and Methods}
(Kluwer Academic Publishers, Boston, 1993).
\end{thebibliography}
\end{document}